\pdfoutput=1
\documentclass[smallextended]{svjour3}       

\usepackage{amsmath,latexsym}
\usepackage{graphicx}
\usepackage{amssymb}
\usepackage[hidelinks]{hyperref}
\usepackage{color}
\usepackage{soul}

\def\L{\mathcal{L}}

\usepackage[a4paper, margin=2.5cm]{geometry}

\begin{document}

\title{Post-Newtonian Hamiltonian dynamics: \\applications to
	stationary    spacetimes  and statistical mechanics }
\titlerunning{Post-Newtonian Hamiltonian dynamics}

\author{Ronaldo S. S. Vieira \and Javier Ramos-Caro \and Alberto Saa}

\institute{R. S. S. Vieira \at
	Centro de Ci\^encias Naturais e Humanas, Universidade Federal do ABC, 09210-580 Santo Andr\'e, SP, Brazil \\
	\email{ronaldo.vieira@ufabc.edu.br}           
	\and
	J. Ramos-Caro \at
    Departamento de F\'\i sica, Universidade Federal de S\~ao Carlos, 13565-905, S\~ao Carlos, SP,
    Brazil \\
    \email{javier@ufscar.br}
	\and
	A. Saa \at
	Departamento de Matemática Aplicada, Universidade Estadual de Campinas, 13083-859 Campinas, SP,
	Brazil \\
	\email{asaa@ime.unicamp.br}
}

\date{Received: date / Accepted: date}

\maketitle

\begin{abstract}
Although the post-Newtonian Lagrangian formalism is widely used in relativistic   dynamical and
statistical  studies 
of test bodies moving around arbitrary mass distributions, the corresponding general Hamiltonian  formalism 
is still relatively uncommon,  being restricted basically to the case of $N$-body problems. Here, 
we  present
a consistent  Hamiltonian formalism  for the 
  dynamics  of test particles in spacetimes with  arbitrary 
energy-momentum distributions in  the first post-Newtonian (1PN) approximation. 
We  apply our formalism to orbital motion in  stationary axisymmetric spacetimes and
 obtain the 1PN 
relativistic corrections to the radial and vertical epicyclic frequencies 
for quasi-circular equatorial motion, a result   potentially interesting for 
  galactic dynamics. For the case of 
 razor-thin disk configurations, 
we obtain  an approximated third integral 
 which could be used to determine  analytically the envelope of nearly equatorial orbits. 
 One of the main advantages of this 1PN analysis is   the explicit presence of 
   frame-dragging effects   in all pertinent expressions, allowing some qualitative predictions in
   rotating spacetimes.   We finish 
 discussing the 1PN collisionless Boltzmann equation in terms of the Hamiltonian
 canonical variables and its relation with previous results in the literature. 
\end{abstract}

\maketitle

\section{Introduction}

The dynamics of test bodies around extended  mass distributions is a fundamental issue in many 
astrophysical scenarios. In the context of weak fields in General Relativity (GR), 
 it is customary to  employ the so-called first post-Newtonian (1PN) approximation
for the description of the motion of test bodies   with relatively low velocities. 
 Apart from the usual Newtonian gravitational  potential $\Phi$, 
the 1PN approximation   includes two additional quantities, the scalar potential 
$\Psi$ and the three-vector $\vec\zeta $, 
representing, respectively, 
  a relativistic correction to the Newtonian potential and  frame-dragging effects.  In this approximation,
   the non-vanishing spacetime metric components  can be written as \cite{weinberg1972gravitation}
  \begin{eqnarray}
  \label{eq:metric1PN--1PN}
    g_{00} &= & -\left(1 + \frac{2\Phi}{c^2} + \frac{2 \Phi^2}{c^4} + \frac{2 \Psi}{c^4}\right), \nonumber \\
    g_{ij} &= & \delta_{ij}\left(1 - \frac{2\Phi}{c^2}\right),\quad 
    g_{0i} =\frac{\zeta_i}{c^3}\,, 
  \end{eqnarray}
  where $c$ stands for the speed of light. 
 Latin indices are assumed to vary from 1 to 3 and correspond to  spatial directions, whereas Greek indices vary from 0 to 3, ranging
 over $(ct,x^i)$. In the  1PN approximation, all analyses and pertinent expressions   are restricted to
 terms up  to $\frac{1}{c^2}$-order. 
 The field equations for $\Phi$, $\Psi$, and $\vec\zeta $ in the 1PN approximation are \cite{weinberg1972gravitation,agonPedrazaRamoscaro2011PRD,ramoscaroAgonPedraza2012PRD}
  \begin{eqnarray}
  \label{eq:fieldequations1PNa}
    \nabla^2\Phi &=& 4\pi G   { \overset{\scriptscriptstyle 0 }{T} }{}^{00}    \,,
        \\
    \nabla^2\Psi - \frac{\partial^2\Phi}{\partial t^2}&=& 4\pi G c^2\left(\overset{\scriptscriptstyle 2}{T}{}^{00} + \sum_i \overset{ \scriptscriptstyle 2 }{T}{}^{ii}\right) ,\\
    \label{eq:fieldequations1PNc}
    \nabla^2\zeta^i &=& 16\pi G c\,\overset{\scriptscriptstyle 1  }{T}{}^{i0}\,, 
  \end{eqnarray}
  where   $\overset{\scriptscriptstyle n \ \ }{T^{\mu\nu}}$ stands for the $\frac{1}{c^n}$-order term
  of the energy-momentum tensor $T^{\mu\nu}$. The differential operators such as $\nabla$ in
  (\ref{eq:fieldequations1PNa})-(\ref{eq:fieldequations1PNc}) are the usual flat-space ones, and indices are raised and
  lowered with the Minkowski metric $\eta_{\mu\nu} = {\rm diag}(-1,1,1,1)$, allowing us to use the usual three-dimensional vector operations
  for the spatial quantities. 
  Furthermore, it is usually assumed in the 1PN approximation the so-called
  harmonic coordinate gauge, $g^{\mu\nu}\Gamma_{\mu\nu}^\delta=0$, which  gives us
    \begin{equation}
   4\frac{\partial\Phi}{\partial t} + \nabla\cdot\vec{\zeta} = 0\,.
  \end{equation}
  The 1PN equations of motion for massive test particles can be  written in terms of their three-velocity
$\vec v$  
   as \cite{weinberg1972gravitation}
\begin{equation}\label{eq:eqmotion}
   \frac{d\vec{v}}{dt} = -\nabla\Big(\Phi + \frac{2\Phi^2}{c^2} + \frac{\Psi}{c^2}\Big)
   + \frac{1}{c^2} \left( \vec{v}\times(\nabla\times\vec{\zeta}\,)
     - \frac{\partial\vec{\zeta}}{\partial t}   
    + \left( {3} \frac{\partial \Phi}{\partial t}  +  {4}(\vec{v}\cdot\nabla)\Phi\right)\vec{v}
  -  {v^2} \nabla\Phi \right),  
\end{equation}
where $v^2 =v_iv^i=|\vec{v}|^2$.
Equation (\ref{eq:eqmotion})  is known to be  derived from the  Lagrangian \cite{weinberg1972gravitation}
\begin{equation}\label{eq:lagrangian1PN}
  \L = \frac{1}{2}v^2 
    + \frac{1}{c^2}\left( \frac{1}{8} v^4  - \frac{3}{2}v^2\Phi     
    + \vec{\zeta}\cdot\vec{v}\right) - \Theta \,,
\end{equation}
where 
  \begin{equation}\label{eq:eta}
\Theta = \Phi + \frac{\Phi^2}{2c^2} + \frac{\Psi}{c^2}\,,
\end{equation}
from which we have the following canonical  momenta 
  \begin{equation}\label{eq:p-i}
   p_i = \frac{\partial \L}{\partial v^i} = v_i\left(1 - \frac{3\Phi}{c^2} + \frac{v^2}{2c^2}\right) + \frac{\zeta_i}{c^2} 
  \end{equation}
and  
energy function
\begin{equation}
\label{eq:energia-em-termos-da-velocidade--1PN}
E = \sum_i p_i v^i - \L =\frac{v^2}{2}  + \frac{1}{c^2}\left(\frac{3v^4}{8}-\frac{3v^2\Phi}{2} \right) + \Theta\,,
\end{equation}
which is indeed conserved for stationary spacetimes   since in this case we will have $\frac{\partial\L }{\partial t} =0$. For further discussions
on the Lagrangian formulation of this problem, see   \cite{agonPedrazaRamoscaro2011PRD,ramoscaroAgonPedraza2012PRD} and references therein.

In the following  sections, we will derive an 1PN Hamiltonian for the Lagrangian (\ref{eq:lagrangian1PN})
and apply it to the
study of orbits in stationary axisymmetric spacetimes.
We first obtain the 1PN relativistic corrections to
the radial and vertical epicyclic frequencies for quasi-circular equatorial motion. Furthermore,
 for the case of razor-thin disk configurations, we get 
an approximated third integral which could be explored to determine analytically the envelope of
nearly equatorial orbits.  We finish discussing the 1PN collisionless Boltzmann equation (CBE) in terms of the
Hamiltonian canonical variables and its relation with some previous results in the literature.

\section{The 1PN Hamiltonian formalism}

The easiest and more direct way of obtaining the Hamiltonian formulation associated with
the Lagrangian (\ref{eq:lagrangian1PN}) is to perform a 1PN Legendre transformation from 
the expression for the canonical momenta 
(\ref{eq:p-i}). Keeping only terms up to $\frac{1}{c^2}$-order, one can invert
 (\ref{eq:p-i}) as 
\begin{equation}\label{eq:vi}
v_i =      p_i\left(1 + \frac{3\Phi}{c^2} - \frac{p^2}{2c^2}\right) - \frac{\zeta_i}{c^2}\,.
\end{equation}
Since the 1PN Hamiltonian $H$ is essentially the function (\ref{eq:energia-em-termos-da-velocidade--1PN}) written
in terms of the canonical variables $(x^i,p_j)$,
we have, keeping again only terms up to $\frac{1}{c^2}$-order, 
  \begin{equation}\label{eq:Heta-1PN}
  H = \frac{1}{2}p^2\left(1 + \frac{3\Phi}{c^2} - \frac{p^2}{4c^2}\right)  
   - \frac{\vec{\zeta}\cdot\vec{p}}{c^2} +  \Theta \,.
  \end{equation} 
  One can check by direct calculations that the 
  Hamilton's equations obtained from (\ref{eq:Heta-1PN}) are equivalent
to the Euler-Lagrange equations of (\ref{eq:lagrangian1PN}). 

We could also have obtained (\ref{eq:Heta-1PN}) by exploring the
so-called isoenergetic reduction formalism \cite{chiconeMashoon2002CQGra}, which basically consists
in considering the  Hamiltonian constraint for timelike geodesics 
$
 g^{\mu\nu}\pi_\mu \pi_\nu  = -c^2,
$
where 
$\pi_\mu = \left(-\frac{E_0}{c},p_i \right)$.
  The reduced Hamiltonian associated with the
time parameter $t = x^0/c$ is   given simply by  $H=E_0 $ \cite{chiconeMashoon2002CQGra}, which can be determined directly 
by solving the quadratic Hamiltonian constraint, leading to
\begin{equation}
\label{iso}
 \frac{ H}{c} =  \sqrt{\frac{c^2 +    \mathfrak{g}^{ij}p_i p_j  }{ -g^{00} }}  +   \frac{ g^{0i}{p_i} }{g^{00}}  
\end{equation}
where
\begin{equation}
\mathfrak{g}^{ij} = g^{i j}-\frac{g^{0 i} g^{0 j}}{g^{00}}\,,
\end{equation}
with the 1PN inverse of (\ref{eq:metric1PN--1PN}) given by \cite{weinberg1972gravitation}
  \begin{eqnarray}  
  \label{contr}
    g^{00} &= & -\left(1 - \frac{2\Phi}{c^2} + \frac{2 \Phi^2}{c^4} - \frac{2\Psi}{c^4}\right), \nonumber \\
    g^{ij} &= & \delta^{ij}\left(1 + \frac{2\Phi}{c^2}\right),\quad 
    g^{0i} =\frac{\zeta^i}{c^3}\,.
  \end{eqnarray}
It is easy to check that (\ref{eq:Heta-1PN}), up to a constant $c^2$ term, follows from (\ref{iso}) by using (\ref{contr}) and keeping only terms up to $\frac{1}{c^2}$-order.
It is worth mentioning that both the 1PN Lagrangian (\ref{eq:lagrangian1PN})  
and    Hamiltonian  (\ref{eq:Heta-1PN}) correspond to the choice $t = x^0/c$ as the
time-evolution parameter for the geodesic dynamics, breaking in this way the GR reparametrization
invariance. Different choices for the time-evolution parameter may result in different Lagrangian
and Hamiltonian 
functions, but nevertheless with fully equivalent dynamics. This can be seen by considering the action for a relativistic particle   
\begin{equation}
\label{lagr}
S = \int 
 {\cal L}\, d\tau , \quad {\cal L}  = -c\sqrt{ -g_{\mu\nu}\dot x^\mu \dot x^\nu } \, ,
\end{equation}
where the dot denotes here derivation with respect to the evolution parameter $\tau$.
Such action is invariant under reparametrizations $\tau \to f(\tau)$, with $\dot{f}(\tau)>0$, and as
a consequence  the canonical momenta
\begin{equation}
\pi_\mu = \frac{\partial \cal L}{\partial \dot x^\mu}  = \frac{c \dot x_\mu}{\sqrt{ - \dot x_\nu \dot x^\nu}}
\end{equation}
obey the Hamiltonian constraint $ \pi_\mu\pi^\mu = -c^2$. The reparameterization-invariant
Lagrangian (\ref{lagr}) is a well-known example of a singular system for which the Hamiltonian
formulation can only be defined properly in the context of constrained systems, see
\cite{Saa} for further details. However, this
problem can be circumvented if we fix a specific parametrization for the action (\ref{lagr}),
breaking in this way the reparametrization invariance of the problem. Choosing 
$\tau = t$, with $x^\mu = (ct,x^i)$, the Lagrangian (\ref{lagr}) reads
\begin{equation}
\label{lagrr}
 {\cal L}  = -c\sqrt{ -\left( c^2g_{00} + 2cg_{0i} \dot x^i + g_{ij}\dot x^i \dot x^j\right) }\,,
\end{equation}
where the dot now denotes derivation with respect to $t$. With the choice $\tau = t$ we have
effectively reduced the dimension of the system, from 4 spacetime dimensions to only 3 spatial ones.
Moreover, the 1PN Lagrangian (\ref{eq:lagrangian1PN}) follows, up to a constant term, from the expansion
of (\ref{lagrr}) up to $\frac{1}{c^2}$-order. For the Hamitonian, 
  we can essentially
follow the same steps of \cite{chiconeMashoon2002CQGra}. The canonical momenta
associated to the reduced Lagrangian (\ref{lagrr}) are
\begin{equation}
\label{pi}
p_i = \frac{\partial \cal L}{\partial\dot x^i} =
 \frac{c^2g_{0i}+2cg_{ij}\dot x^j}{\sqrt{ -\left( c^2g_{00} + 2cg_{0i} \dot x^i + g_{ij}\dot x^i \dot x^j\right) }} 
\end{equation}
and the associated Hamiltonian function will be
\begin{equation}
{\cal H} = p_i\dot x^i - {\cal L} =
-c \frac{c^2 g_{00} + cg_{0i}\dot x^i}{\sqrt{ -\left( c^2g_{00} + 2cg_{0i} \dot x^i + g_{ij}\dot x^i \dot x^j\right) }} = -c \pi_0\,,
\end{equation}
where the expressions for the momenta (\ref{pi}) were used in the last equation. We can now solve the Hamiltonian constraint
$ \pi_\mu\pi^\mu = -c^2$ and write $\pi_0$ in terms of the spatial momenta, and essentially we
reproduce the isoenergetic
reduction construction  and get (\ref{iso}). 
The freedom in choosing    different evolution parameters 
 will be important in order to properly compare  different Hamiltonian and Lagrangian formulations  of the CBE
in the following sections.

\section{Stationary, axisymmetric configurations}

Let us start by applying  the  1PN Hamiltonian formalism to the dynamics of
 test particles in stationary, axially symmetric spacetimes. 
We will use the usual
  cylindrical coordinates $(ct,r, \varphi,z)$ and assume reflection symmetry with respect to the equatorial
plane $z=0$. The corresponding gravitational potentials
  are independent of $t$ and $\varphi$ and, moreover,  have well-defined 
  parity properties 
  with respect to $z\to -z$ reflections. 
In cylindrical coordinates, we have 
\begin{equation}\label{eq:p2axisymmetric}
   p^2 =   p_r^2  + p_z^2 + \frac{L_z^2}{r^2}\,,
\end{equation}
where we have already used that   $\varphi$ is a cyclic coordinate and hence $p_\varphi=L_z$ is a constant of motion.
Since we are dealing with stationary and axially symmetric spacetimes, the post-Newtonian metric must be a special case of the 
Weyl-Lewis-Papapetrou metric, and therefore we must have $g_ {0r} = g_{0z} =0$.
The rotation term is, thus, purely centrifugal. Therefore, our vector potential must have the form
$
   \vec{\zeta} = \zeta_{\hat\varphi}\,\hat\varphi,
$
where $\zeta_{\hat\varphi}$ is the azimuthal component of the vector $\vec{\zeta}$ in the
orthonormal frame $(\hat r, \hat \varphi,\hat z)$. The frame-dragging term can be written as
\begin{equation}
   \vec{\zeta}\cdot\vec{p} = \frac{\zeta_{\hat\varphi} L_z}{r}\,.
\end{equation}
Introducing the quantities 
\begin{eqnarray}
   A(r,z) &=& 1 + \frac{1}{c^2}\left(3\Phi - \frac{L_z^2}{2r^2}\right), \label{f} \\
   B(p_r,p_z,r,z) &=& A(r,z) -\frac{1}{4c^2}(p_r^2 + p_z^2)\,, \label{ftilde}\\
   C(r,z) &=& 1 + \frac{1}{c^2}\Big(3\Phi - \frac{L_z^2}{4r^2}\Big),
\end{eqnarray}
and the effective potential
\begin{equation}
\label{eq:Veffstationary}
	   V_{\rm{eff}} = \Theta + C(r,z)\frac{L_z^2}{2r^2} - \frac{1}{c^2}\frac{\zeta_{\hat\varphi} L_z}{r}\,, 
\end{equation}
we can write the Hamiltonian (\ref{eq:Heta-1PN}) simply as
  \begin{equation}\label{eq:Hstationary}
   H = \frac{1}{2}  \big(p_r^2 + p_z^2\big)B(p_r,p_z,r,z) + V_{\rm{eff}}(r,z)\,.
  \end{equation}
 Notice that the
positiveness of the    kinetic term  requires  $B>0$, which is indeed verified if we consider
the $\frac{1}{c^2}$-order terms as subdominant. 

\subsection{Quasi-circular motion}

Circular orbits are given by the fixed points of the Hamiltonian when restricted to the equatorial plane $z=0$. They are given by the conditions $p_r = p_z = 0$ and are the solutions for $r$ of
  \begin{equation}
  \label{circularorbitVeffr}
   \frac{\partial V_{\rm{eff}}}{\partial r}= \frac{\partial \Theta}{\partial r} - 
   \frac{L_z^2}{r^3}  +   \frac{1}{c^2}\left[  
  \left(\frac{3r}{2 }\frac{\partial\Phi}{\partial r}  - {3\Phi} + \frac{ L_z^2}{2 r^2}\right)\frac{L_z^2}{r^3}  -  {L_z}\frac{\partial}{\partial r}\bigg(\frac{\zeta_{\hat\varphi}}{r}\bigg)\right] = 0
  \end{equation}  
  at $z=0$. 
 In order to determine the angular momentum $L_z$ of a circular orbit, we 
solve (\ref{circularorbitVeffr})  
  explicitly for $L_z$ up to $\frac{1}{c^2}$-order, obtaining 
  \begin{equation}
  \label{ang}
  L_z^2(r) = \left( L_z^N\right)^2  \left[ 1 
  -  \frac{2}{c^2}\left( \Phi - \frac{\left( L_z^N\right)^2}{r^2}\right) \right]
  +\frac{r^3}{c^2}\left( \frac{\partial \Psi}{\partial r} -  L_z^N \frac{\partial}{\partial r}\Big(\frac{\zeta_{\hat\varphi}}{r}\Big)\right),
  \end{equation}
where $L_z^N$ is the corresponding Newtonian angular momentum, which is known to be  given by
\begin{equation}
  L_z^N   = \pm \sqrt{ r^3\frac{\partial\Phi}{\partial r}}\,.
\end{equation}
The $L_z^N$ linear contribution
in   the last term of (\ref{ang})  explicitly exhibits  a frame-dragging effect.
The 1PN azimuthal angular momentum $L_z^2(r)$ for circular orbits, Equation  (\ref{ang}), corresponds to a circular velocity $\vec v = v_{\hat{\varphi}}\,\hat{\varphi}$, where   $v_{\hat{\varphi}} = r\dot{\varphi}$, which can be determined from (\ref{eq:vi}), leading to
\begin{equation}\label{eq:vphi2}
	v_{\hat{\varphi}}^2 = r\frac{\partial\Phi}{\partial r}
 + \frac{r}{c^2}\left[4\Phi\frac{\partial\Phi}{\partial r} + \frac{\partial\Psi}{\partial r} + r\Big(\frac{\partial\Phi}{\partial r}\Big)^2 - \frac{L_z^N}{r}\,\Big(\frac{\partial\zeta_{\hat\varphi}}{\partial r}+\frac{\zeta_{\hat\varphi}}{r}\Big)\right].
 \end{equation}
Notice that the second term   in the frame-dragging contribution  is lacking in the equation (A5)
of \cite{ramoscaroAgonPedraza2012PRD}.
 
We now turn our attention to the radial and vertical epicyclic frequencies of quasi-circular motion.
Small perturbations to the circular orbit are expected to obey   linear
  equations around the fixed points, {\em i.e.},
  \begin{equation}
   \delta\ddot r = - \kappa^2 \delta r\,, \quad
   \delta\ddot z = -  \nu^2 \delta z\,,
  \end{equation}
  where $\kappa$ and $\nu$ are, respectively, the radial and vertical 
epicyclic frequencies and in our case are given by
  \begin{eqnarray}
   \kappa^2  &=& A(r,0)\frac{\partial^2 V_{\rm{eff}}}{\partial r^2} =
    \kappa^2_N + \kappa^2_{\rm 1PN}\,, \label{defkappa2} \\
   \nu^2  &=& A(r,0) \frac{\partial^2 V_{\rm{eff}}}{\partial z^2} =
   \nu^2_N + \nu^2_{\rm 1PN}\, , \label{defnu2}
  \end{eqnarray}
     evaluated at $L_z = L_z(r)$, Equation (\ref{ang}).
 Here,  $\kappa_N$ and $\nu_N$ stand for,  respectively, the radial and vertical  Newtonian
epicyclic frequencies 
  \begin{eqnarray}
   \kappa^2_N &=& \left(\frac{\partial^2\Phi}{\partial r^2} + 
  \frac{3}{r}\frac{\partial\Phi}{\partial r}\right), \label{kappa2newt}\\
   \nu^2_N &=& \frac{\partial^2\Phi}{\partial z^2} \,. \label{nu2newt}
  \end{eqnarray}
The 1PN corrections for the   epicyclic frequencies are
  \begin{equation}
   \kappa^2_{\rm 1PN} = \frac{1}{c^2}\left\{\,  4\Phi\kappa^2_N + 
 \Big(\frac{\partial^2\Psi}{\partial r^2} + \frac{3}{r}\frac{\partial\Psi}{\partial r}\Big)   + 
 \frac{\left( L_z^N\right)^2}{r^2} \Big(\frac{\partial^2\Phi}{\partial r^2}  - 
 \frac{3}{r}\frac{\partial\Phi}{\partial r}\Big)   - L_z^N\left[\frac{\partial^2}{\partial r^2}\Big(\frac{\zeta_{\hat\varphi}}{r}\Big)
  + \frac{3}{r}\frac{\partial}{\partial r}\Big(\frac{\zeta_{\hat\varphi}}{r}\Big)\right]    \,\right\}
  \label{kappa2stationary} 
\end{equation} 
  and  
  \begin{equation}
  \nu^2_{\rm 1PN} =\frac{1}{c^2}\left\{\,   4\Phi \nu^2_N +\frac{\partial^2\Psi}{\partial z^2} +
  \Big(\frac{\partial\Phi}{\partial z}\Big)^2  + 
  \frac{\left( L_z^N\right)^2}{r^2}  
  \frac{\partial^2\Phi}{\partial z^2}  
  - \frac{L^N_z}{  r} \frac{\partial^2\zeta_{\hat\varphi}}{\partial z^2} \,\right\} . \label{nu2stationary}
  \end{equation}
  One can appreciate again the frame-dragging effect in both radial and
  vertical 
   epicy\-clic frequencies for rotating and counter-rotating equatorial circular   orbits. 
   All the above expressions may be relevant for post-Newtonian corrections to galactic dynamics. Equation~(\ref{eq:vphi2}) gives us the 1PN corrections to the rotation curves of spiral and elliptical galaxies, including the frame-dragging term, and Equations~(\ref{kappa2stationary}) and (\ref{nu2stationary}) give us the 1PN corrections to the epicyclic frequencies of quasi-circular motion, having impact for instance on the radii of Lindblad resonances (the angular speed is given by $\Omega(r) = v_{\hat{\varphi}}/r$). Moreover, the Hamiltonian (\ref{eq:Hstationary}) extends the corresponding Newtonian expression and may be useful to the study of chaos in 1PN configurations. These topics, however, are beyond the scope of the present work.

\subsection{Approximate third integral of motion for razor-thin disks}

We now consider razor-thin disks in the presence of the scalar potential $\Psi$ and the vector potential $\vec\zeta = \zeta_{\hat\varphi}\,\hat\varphi$, a situation corresponding to the inclusion of $\frac{1}{c^2}$--order
frame-dragging effects.
From the field equations (\ref{eq:fieldequations1PNa})--(\ref{eq:fieldequations1PNc}), 
we see that 
the vector potential generates an azimuthal momentum flux,  but it does not affect neither the energy density nor the principal pressures of the disk. Since the 1PN field equations (\ref{eq:fieldequations1PNa})--(\ref{eq:fieldequations1PNc}) are linear in the fields $\Phi$, $\Psi$ 
and $\vec\zeta$, we can linearly superpose their solutions in order to consider more general systems. 
In particular, we can superpose the razor-thin disk to axially symmetric structures such as a thick disk, a spheroidal bulge
and a spheroidal (stellar + dark matter) halo, as in mass models for spiral galaxies \cite{allenSantillan1991RMxAA,helmi2004MNRAS,mcmillan2011MNRAS,vieiraRamoscaro2014ApJ,vieiraRamoscaro2019MNRAS}. Anyway, we can always analyze 
the thin-disk contribution separately from the contribution due to the surrounding matter.

The energy-momentum tensor  for a razor-thin disk  
has the form $T^{\mu\nu}\propto\delta(z)$, see for instance \cite{vieiraramosCaro2016CeMDA,vieiraramoscaroSaa2016PrD,vieira2020CQGra} and references therein. 
We can write it as \cite{agonPedrazaRamoscaro2011PRD,ramoscaroAgonPedraza2012PRD} 
  \begin{eqnarray}
 \label{Tmunuthinstationary}
    \overset{\scriptscriptstyle n \ \ }{T^{00}} &=& \overset{n}{\sigma}(r)\delta(z) \\
     \overset{\scriptscriptstyle 2 \ \ }{T^{ii}} &=& \frac{\overset{\scriptscriptstyle 2}{P}_i(r)}{c^2}\delta(z)    \\
    \overset{\scriptscriptstyle 1 \ \ }{T^{0\varphi}} &=&  \frac{\mu(r)}{c}\delta(z)\,,   
  \end{eqnarray}
where $\sigma = \overset{\scriptscriptstyle 0}{\sigma} + \overset{\scriptscriptstyle 2}{\sigma}$ is the surface mass density,
 $\overset{\scriptscriptstyle 2}{P}_i$ are the principal 
pressures, and  $\mu(r)$ is the azimuthal
surface momentum density of the disk. We can now evaluate the field equations (\ref{eq:fieldequations1PNa})--(\ref{eq:fieldequations1PNc}) near the equatorial plane $z=0$ and obtain 
  \begin{eqnarray}
  \label{cc1}
    \frac{\partial\Phi}{\partial |z|} &=& 2\pi G\overset{\scriptscriptstyle 0}{\sigma},\\
    \frac{1}{c^2}\frac{\partial\Psi}{\partial |z|} &=& 2\pi G \Big(\overset{\scriptscriptstyle 2}{\sigma} +  \frac{\mathcal{P}}{c^2}\Big),\\
    \label{cc2}
   \frac{1}{r}\frac{\partial\zeta_{\hat\varphi}}{\partial |z|} &=& 8\pi G  {\mu} \,,
  \end{eqnarray}
  where 
$
\mathcal{P} = P_r + P_\varphi
$ since razor-thin disks have no vertical pressure. 
Let us now consider the vertical stability of equatorial circular orbits in  1PN stationary razor-thin disks.
From (\ref{eq:Veffstationary}) and (\ref{cc1})--(\ref{cc2}), we have
  \begin{equation}\label{partialVeffpartialabszstationary}
  \frac{1}{ 2\pi G} \frac{\partial V_{\rm{eff}}}{\partial |z|} =
  \overset{0}{\sigma} + \overset{2}{\sigma} + \frac{1}{c^2}\left( \mathcal{P}  -4\frac{ L_z\mu}{r}
    + \Big(\Phi + 
  \frac{3L_z^2}{2r^2}\Big) \overset{0}{\sigma}\right)  .
  \end{equation}
The  usual vertical stability criterion for   a circular orbit is $\frac{\partial V_{\rm{eff}}}{\partial |z|}>0$ at $z=0$ \cite{vieiraramosCaro2016CeMDA,vieiraramoscaroSaa2016PrD}, implying  that,
since  $ \overset{\scriptscriptstyle 0}{\sigma}$ dominates the matter contributions, 
the Newtonian condition $ \overset{\scriptscriptstyle 0}{\sigma}>0$ is enough to guarantee vertical
stability of equatorial circular orbits.

In order to obtain an approximate third integral of motion for nearly equatorial   orbits (in addition to $E$ and $L_z$), we follow closely  \cite{vieiraramosCaro2016CeMDA,vieiraramoscaroSaa2016PrD}. 
The approximate Hamiltonian $   H^{(a)}$ for the motion
 near the stable equatorial circular orbit of radius $r_o$ is written as
  \begin{equation}
   H^{(a)} = \frac{A(r_o,0)}{2}\Big[p_r^2 + p_z^2\Big] + V_{\rm{eff}}^{(a)},
  \end{equation}
where the approximate effective potential is given by
  \begin{equation}
   V_{\rm{eff}}^{(a)} = V_{\rm{eff}}(r,0) +\omega_{r_o}|z|
  \end{equation}
  with
  \begin{equation}
   \omega_{r_o} = \frac{\partial V_{\rm{eff}}}{\partial |z|}(r_o,0)\,.
  \end{equation}  
  One can now separate the Hamiltonian as $H^{(a)} = H_r + H_z$,
where 
  \begin{equation}
   H_z = \frac{{A(r_o,0)}}{2}p_z^2 + \omega_{r_o}|z|\,.
  \end{equation}
  The corresponding action variable can be written as \cite{vieiraramosCaro2016CeMDA,vieiraramoscaroSaa2016PrD}
  \begin{equation}
   J_z = \frac{4}{3\pi}\sqrt{\frac{2\,\omega_{r_o}Z^3}{{A(r_o,0)}}} \, ,
  \end{equation}
where $Z$ is the vertical amplitude of the motion. Considering that $r_o$ varies slowly in
time due to off-equatorial motion, one can explore adiabatic invariance arguments for the 
action variable $J_z$ and derive a relation between $Z$ and $r$, establishing in this
way the orbit’s envelope. Introducing the quantity 
  \begin{equation}\label{tildeomega}
   \tilde\omega(r) = \frac{1}{A(r,0)}\frac{\partial V_{\rm{eff}}}{\partial |z|}(r,0)
  \end{equation}
  (see Equation (\ref{partialVeffpartialabszstationary})), we have that $  \tilde\omega(r) Z^3$ is approximately   a constant, which gives us the envelopes for off-equatorial orbits
  \begin{equation}
  	Z(r)\propto \big[\tilde\omega(r)\big]^{-1/3}\,.
  \end{equation}
In terms of the original canonical variables we can write 
  \begin{equation}
I_3= \big[\tilde\omega(r)\big]^{-2/3}\left(\,\frac{1}{2}p_z^2 + \tilde\omega(r)|z|\,\right).
\end{equation}
The quantity $I_3$ is therefore an approximate third integral of motion for our problem. The novelty here, with respect
to the Newtonian \cite{vieiraramosCaro2016CeMDA} and the relativistic \cite{vieiraramoscaroSaa2016PrD} cases, 
 is that this integral
and, consequently, the envelope of the orbits, exhibit explicitly the frame-dragging effect arising
from the $L_z$ linear term in  (\ref{partialVeffpartialabszstationary}) and, consequently,
in $\tilde\omega(r)$.

\section{Post-Newtonian statistical mechanics}

The distribution function  $f(x^i, p_j,t)$ of a system of particles   whose motion is described by the Hamiltonian flow of $H$, in the absence
of collisions, satisfies the so-called  collisionless Boltzmann equation (CBE)
  \begin{equation}\label{eq:CBE} 
   \frac{df}{dt} = \frac{\partial f}{\partial t} + \{f, H\} = 0\,,
  \end{equation}
 with $\{\ , \  \} $ standing for the usual Poisson brackets, where $(x^i, p_j)$ are canonical coordinates. In our case, the CBE reads
   \begin{eqnarray}\label{eq:CBEf}
   \frac{df}{dt}  &=& \frac{\partial f}{\partial t} + p^i\frac{\partial f}{\partial x^i} - 
    \frac{\partial \Phi}{\partial x^i}\frac{\partial f}{\partial p_i} + \frac{1}{c^2}  \left(\left(3\Phi - \frac{p^2}{2}\right)p^i - \zeta^i\right)\frac{\partial f}{\partial x^i}   \nonumber \\
   && - \frac{1}{c^2} \left(\left(\Phi + \frac{3}{2}p^2\right)\frac{\partial \Phi}{\partial x^i} + \frac{\partial \Psi}{\partial x^i}
    - {p}_j\frac{\partial  {\zeta^j}}{\partial x^i}\right)\frac{\partial f}{\partial p_i} = 0\,,
  \end{eqnarray}
  where summation over repeated indices is assumed.
The formulation of the CBE in terms of canonical momenta  may be useful for constructing self-consistent, stationary models of self-gravitating systems in the 1PN approximation  as a complementary tool to the results developed in \cite{agonPedrazaRamoscaro2011PRD,ramoscaroAgonPedraza2012PRD}. It also gives us, in a direct way, the conserved quantities associated with symmetries of the dynamics, for instance energy in stationary spacetimes and angular momentum in spherically symmetric spacetimes. The expressions obtained are the same of \cite{agonPedrazaRamoscaro2011PRD}. 

Nevertheless, for sake of comparison with some previous results in the literature, let us consider the distribution function $\tilde f$ in the Lagrangian formulation, $ f(x^i, p_j,t) = \tilde f(x^i,v^j(x,p,t),t)$. We have, up
to $\frac{1}{c^2}$ order, 
\begin{eqnarray}
\frac{\partial f}{\partial x^i} &=&   \frac{\partial \tilde f}{\partial x^i} + \frac{1}{c^2}
\left( 3 v^j \frac{\partial \Phi}{\partial x^i} - \frac{\partial \zeta^j}{\partial x^i}\right)
\frac{\partial \tilde f}{\partial v^j} \\ 
\frac{\partial f}{\partial p_i} &=&   \frac{\partial \tilde f}{\partial v_i} + \frac{1}{c^2}
\left[  \left( 3\Phi - \frac{v^2}{2} \right)\frac{\partial \tilde f}{\partial v_i} - v^iv^j\frac{\partial \tilde f}{\partial v^j}\right]
 \\ 
\frac{\partial f}{\partial t} &=&   \frac{\partial \tilde f}{\partial t} + \frac{1}{c^2}
\left(  
3v^j \frac{\partial\Phi}{\partial t}-\frac{\partial\zeta^j}{\partial t}
\right)
\frac{\partial \tilde f}{\partial v^j}  
\end{eqnarray}
giving from (\ref{eq:CBEf}) the following equation for the distribution function $\tilde f$,
\begin{equation} 
\label{eq:CBEL}
\frac{\partial\tilde f}{\partial t} + v^i\frac{\partial \tilde f}{\partial x^i}  + \dot v^i  \frac{\partial \tilde f}{\partial v^i}  =0\,,
\end{equation}
with $\dot{v}^i$ given by (\ref{eq:eqmotion}), confirming again the consistency of our results. 
Equation~(\ref{eq:CBEL2}) can be compared with the 1PN Liouville equation presented in \cite{agonPedrazaRamoscaro2011PRD}, their equation (9). For a better comparison,
let us write (\ref{eq:CBEL}) explicitly with all the spatial terms,
\begin{equation}
	 \frac{\partial \tilde f}{\partial t} + v^i\frac{\partial \tilde f}{\partial x^i} - \frac{\partial\Theta}{\partial x^i}\frac{\partial \tilde f}{\partial v_i} + \frac{1}{c^2}\bigg\{4 v_i v^j \frac{\partial\Phi}{\partial x^j} - v^2 \frac{\partial\Phi}{\partial x^i} + 
	3 v_i\frac{\partial\Phi}{\partial t}   \label{eq:CBEL2}+ v^j\left(\frac{\partial \zeta_j}{\partial x^i} - \frac{\partial \zeta_i}{\partial x^j}\right) - \frac{\partial \zeta_i}{\partial t}
	\bigg\}\,\frac{\partial \tilde f}{\partial v_i} = 0 \,. 
\end{equation}
Although they seem to be different at first look, one may check that the difference arises because of a global multiplying factor $U^0/c = dt/d\tau$ in their equation, where $\tau$ is the proper time along the trajectory. More
specifically, 
 writing Eq.~(\ref{eq:CBEL2}) as $\mathcal{L}_v f = 0$, where $\mathcal{L}_v$ is the Liouville operator, it follows that the Liouville operator $\mathcal{L}_U$ of \cite{agonPedrazaRamoscaro2011PRD} corresponds to $\mathcal{L}_U = (U^0/c)\,\mathcal{L}_v$.  The equations are dynamically equivalent; their difference in form can be
attributed to different choices for the time-evolution parameter. 
One advantage of the present formulation is that the 1PN correction to the Newtonian case appears only in the term $\dot v^i\,\frac{\partial \tilde f}{\partial v^i}$, showing explicitly the post-Newtonian correction in the Lagrangian formulation of the CBE.

\section{Conclusions}

We presented here, for the first time, the Hamiltonian formalism  for the dynamics of test particles in spacetimes with  arbitrary 
energy-momentum distributions in  the first post-Newtonian (1PN)  approximation. This construction is consistent with the corresponding  Lagrangian formalism up to 1PN order and therefore gives an alternative way to deal with particle dynamics in 1PN self-gravitating systems, with the advantage of presenting explicitly the symplectic nature of the dynamics.

The expressions obtained for physical observables in  (stationary) axially symmetric configurations, such as the 1PN circular speed and (radial and vertical) epicyclic frequencies of quasi-circular equatorial motion in the presence of frame-dragging effects, are useful for further studies including 1PN terms in galactic dynamics, giving us relativistic corrections to rotation curves of spiral and elliptical galaxies, and to the positions of Lindblad resonances. Moreover, the stationary, axially symmetric Hamiltonian (\ref{eq:Hstationary}), or even the general expression (\ref{eq:Heta-1PN}), permits us to perform systematic studies of chaos in 1PN self-gravitating systems. One application of this formalism to the dynamics of disk-crossing orbits was presented here, the  approximated third integral of motion for stationary, axially symmetric razor-thin disk systems 
which also determines  analytically the envelope $Z(r)$ of nearly equatorial orbits. We emphasize that in all our results we could identify  the explicit presence of frame-dragging effects.  

We also obtained the 1PN collisionless Boltzmann equation in terms of the canonical variables ($x^i, p_j$), and of the Hamiltonian (\ref{eq:Heta-1PN}). This form of the CBE may be useful for constructing self-gravitating models for the distribution function of gravitating stellar systems in terms of canonical momenta and of action variables when GR corrections are taken into account, as a complementary method to the results of \cite{agonPedrazaRamoscaro2011PRD,ramoscaroAgonPedraza2012PRD} based on the Lagrangian formalism. We point out that the CBE~(\ref{eq:CBEf}), when written in velocity space ($x^i, v^j$), Equation~(\ref{eq:CBEL2}), gives a different result from \cite{agonPedrazaRamoscaro2011PRD}. This difference, however, is due solely to a global multiplying factor in the Liouville operator, giving dynamically equivalent results. One advantage of our formalism is that, when written in terms of ($x^i, v^j$), the post-Newtonian corrections appear only in the $\partial\tilde f/\partial v^i$ term, as expected from the 1PN corrections to the equations of motion in the Lagrangian formalism.

One potentially interesting application of the results presented here is the construction of 1PN spherical self-consistent models and of distribution functions for 1PN razor-thin disks, possibly in a more direct way than in the Lagrangian formalism \cite{agonPedrazaRamoscaro2011PRD,ramoscaroAgonPedraza2012PRD}, helping to extend the results presented in those works.

 \section*{Acknowledgements}
 The work of AS is 
 partially supported by CNPq,  grant  302674/2018-7.


\begin{thebibliography}{10}
	\providecommand{\url}[1]{{#1}}
	\providecommand{\urlprefix}{URL }
	\expandafter\ifx\csname urlstyle\endcsname\relax
	\providecommand{\doi}[1]{DOI \discretionary{}{}{}#1}\else
	\providecommand{\doi}{DOI \discretionary{}{}{}\begingroup
		\urlstyle{rm}\Url}\fi
	
	\bibitem{weinberg1972gravitation}
	S. Weinberg: Gravitation and cosmology: Principles and applications of
	general theory of relativity. John Wiley and Sons, Inc., New York (1972).
	
	\bibitem{agonPedrazaRamoscaro2011PRD}
	C.A. {Ag{\'o}n}, J.F. {Pedraza}, J.~{Ramos-Caro}: Kinetic theory of collisionless self-gravitating gases: Post-Newtonian polytropes. Phys. Rev. D \textbf{83}, 123007
	(2011).
	\newblock \url{https://doi.org/10.1103/PhysRevD.83.123007}
	
	\bibitem{ramoscaroAgonPedraza2012PRD}
	J.~{Ramos-Caro}, C.A. {Ag{\'o}n}, J.F. {Pedraza}: Kinetic theory of collisionless self-gravitating gases. II. Relativistic corrections in galactic dynamics. Phys. Rev. D \textbf{86}, 043008
	(2012).
	\newblock \url{https://doi.org/10.1103/PhysRevD.86.043008}
	
	\bibitem{chiconeMashoon2002CQGra}
	C.~{Chicone}, B.~{Mashhoon}: The generalized Jacobi equation. Class. Quant. Grav. \textbf{19}, 4231
	(2002).
	\newblock \url{https://doi.org/10.1088/0264-9381/19/16/301}
	
	\bibitem{Saa} A. Saa:  Canonical quantization of the relativistic particle in static spacetimes.
	Class. Quant. Grav. {\bf 13}, 553 (1996).	\newblock \url{https://doi.org/10.1088/0264-9381/13/3/018}	
	
	\bibitem{allenSantillan1991RMxAA}
	C.~{Allen}, A.~{Santill{\'a}n}: An improved model of the galactic mass distribution for orbit computations. Rev. Mexicana Astron. Astrofis. \textbf{22}, 255 (1991).
	\newblock \url{http://adsabs.harvard.edu/abs/1991RMxAA..22..255A}
	
	\bibitem{helmi2004MNRAS}
	A.~{Helmi}. Is the dark halo of our Galaxy spherical?. Mon. Not. R. Astron. Soc. \textbf{351}, 643 (2004).
	\newblock \url{https://doi.org/10.1111/j.1365-2966.2004.07812.x}.
	
	\bibitem{mcmillan2011MNRAS}
	P.J. {McMillan}. Mass models of the Milky Way. Mon. Not. R. Astron. Soc. \textbf{414}, 2446 (2011).
	\newblock \url{https://doi.org/10.1111/j.1365-2966.2011.18564.x}
	
	\bibitem{vieiraRamoscaro2014ApJ}
	R.S.S. {Vieira}, J.~{Ramos-Caro}: A Simple Formula for the Third Integral of Motion of Disk-Crossing Stars in the Galaxy. Astrophys. J. \textbf{786}, 27 (2014).
	\newblock \url{https://doi.org/10.1088/0004-637X/786/1/27}
	
	\bibitem{vieiraRamoscaro2019MNRAS}
	R.S.S. {Vieira}, J.~{Ramos-Caro}: Envelopes and vertical amplitudes of disc-crossing orbits. Mon. Not. R. Astron. Soc. \textbf{484}, 5155 (2019).
	\newblock \url{https://doi.org/10.1093/mnras/stz325}
	
	\bibitem{vieiraramosCaro2016CeMDA}
	R.S.S. Vieira, J.~Ramos-Caro: Integrability of motion around galactic razor-thin disks. Celestial Mechanics and Dynamical Astronomy
	\textbf{126}(4), 483 (2016).
	\newblock \url{https://doi.org/s10569-016-9705-0}
	
	\bibitem{vieiraramoscaroSaa2016PrD}
	R.S.S. Vieira, J.~Ramos-Caro, A.~Saa: Vertical stability of circular orbits in relativistic razor-thin disks. Phys. Rev. D \textbf{94}, 104016 (2016).
	\newblock \url{https://doi.org/10.1103/PhysRevD.94.104016}
	
	\bibitem{vieira2020CQGra}
	R.S.S. Vieira: Self-gravitating razor-thin discs around black holes via multi-hole seeds. Class. Quant. Grav. \textbf{37}, 205013, (2020).
	\newblock \url{https://doi.org/10.1088/1361-6382/aba99b}
	
	
	
\end{thebibliography}

\end{document}